\documentclass[aps]{revtex4}
\usepackage{graphicx}

\def\bea{\begin{eqnarray}}
\def\eea{\end{eqnarray}}
\def\ket#1{\vert #1 \rangle}

\def\sqr#1#2{{\vcenter{\vbox{\hrule height.#2pt
      \hbox{\vrule width.#2pt height#1pt \kern#1pt
         \vrule width.#2pt}
      \hrule height.#2pt}}}}

\def\figloc#1#2 {
\begin{figure}\begin{center}
    \includegraphics[width=80mm]{fig#1.ps}
    \caption{Figure #1. #2}
    \end{center}\end{figure}
}

\begin{document}
\title{ Decoherence without Dissipation   }

\author{W. G. Unruh}
\affiliation{ CIAR Cosmology and Gravity Program\\
Dept. of Physics\\
University of B. C.\\
Vancouver, Canada V6T 1Z1\\
~
email: unruh@physics.ubc.ca}

~

~

\begin{abstract}
That decoherence can take place in the presence of energy conservation seems
to be a poorly known fact. That lack of knowledge has for example bedevilled
the discussion of the ``black hole information" problem. I present a simple
model which illustrates such energy free decoherence. 

\end{abstract}

\maketitle

It has become a truism in the discussions of the quantum behaviour of black
holes that if black holes destroy the coherence of the quantum state outside
the black hole, then that must lead to energy non-conservation. This argument
was first expressed in the paper by Banks et al\cite{banks} and has become a
``fact" leading to the regard of the so called black hole information paradox
as one of the greatest problems facing quantum gravity, despite the arguments
of Unruh and Wald\cite{wald} that those arguments were suspect. In short,
black holes remember the amount of energy that went into their formation in
the gravitational field surrounding the black hole. Mass, angular momentum and
charge are all encoded in the fields surrounding the black hole, and as such
can and will influence the radiation emitted via the Hawking process in black
hole evaporation. 

In this paper, I want to look at a very simple problem, the scattering of two
particles off each other, in which, if one examines only the two particles,
one finds that the scattering process preserves the energy after the
scattering event (ie energy before equals energy afterwards) but the density
matrix for the state of the two particles goes from a pure state to a mixed
state after the scattering. 

This demonstration will take place, not in the context of quantum gravity, but
of ordinary quantum mechanics, in which case the only way that conversion can
take place is by the system becoming correlated (entangled)  with another
``hidden" quantum system. In the black hole case, this is the "singularity"
within the black hole which could equally well be regarded with no loss of
generality as some sort of baby universe, or eventually disconnected
spacetime. 

In my model, I will use a ``spin bath" as the hidden sector. By "spins" I mean
any
two level quantum systems, which can always be modelled as if it were a three
dimensions spin 1/2 system. In what follows I will refer to these as spins
(without the quotes) although they should not necessarily be seen as real
spins, but as generic two-dimensional-Hilbert-space objects. My model is
closely related to the ``spin decoherence" emphasized by Stamp and
Prokof'ev\cite{stamp}
as a dominant decoherence mechanism in some condensed matter systems. (In
their case the ``spins" are often real spins, but can often be any system with
a finite dimensional Hilbert space.) They also
found that their ideas met with initial disbelief in the condensed matter community,
although the ideas have had experimental confirmation\cite{stamp-exp}
Their ideas have still not diffused into general knowledge in their commmunity, just as Unruh
and Wald's ideas have still resisted diffusion in the gravitational/particle physics community.

Consider two particles, which for sake of simplicity I will assume have the
same mass, and live in a 1+1 dimensional spacetime. They interact only when in
contact with each other, and their interaction is mediated by some hidden
degrees of freedom which are represented by a number $N$ of spin 1/2 objects, with spins
operators $\vec S_i$. The interaction Hamiltonian is assumed to be of the
form $\delta(x_1-x_2) \sum_i S^3_i$
where $S^3={1\over 2}\sigma^3$ the third Pauli spin matrix, while the Kinetic
energy is the usual ${1\over 2m}(p_1^2.+p_2^2)$. 

The spins alter the interaction potential of the two particles, making that
potential a function of the spins. At the same
time, the two objects when located together, cause the spins to precess around
the $z$ axis. It is the correlation introduced between the scattering of the
particles and the dirction of the spin vector which leads to the entanglement
between the two system, and the decoherence of the particle scattering.

Going into the  centre of mass coordinates
\bea
Y&=&(x_1+x_2)/2 \\
y&=&(x_1-x_2)
\eea
the Schroedinger equation  becomes
\bea
i\partial_t \Psi(t,Y,y,\{s_i\}) = -{1\over m}\partial_Y^2 \Psi -{1\over
2m}\partial_y^2 \Psi +\mu\delta(y)(\sum_i S^3_i)\Psi 
\eea
where $\mu$ is a coupling constant.

This system has interactions which are time independent, and thus the energy
is an exactly conserved quantity, and one can find the energy eigenstates of
the system.
The energy eigenstates of this model can be solved by separation of variables
\bea
\Psi_{E+\epsilon} =\psi_E(Y)\phi_\epsilon(y,\{s^3_i\})  
\eea
where the $s^3_i$ are the eigenvalues of $S^3_i$.
and 
\bea
-\partial_Y^2\psi&=&mE\psi\\
-{1\over 2m}\partial_y^2\phi +\mu\delta(y)(\sum_is^3_i) &=&\epsilon \phi
\eea
and the total energy is $E+\epsilon$.

The first equation has the usual plane wave states,
$e^{iKY}$,  as
solutions with $K=\pm\sqrt{mE}$. The second has the scattering states
\bea
\phi_{\epsilon,\{s^3_i\},+}= e^{iky} +A_{k,s_i}e^{-iky}\Theta(-y) +
B_{k,s_i}e^{iky}\Theta(y)\\
\phi_{\epsilon,\{s^3_i\},-}=\left( e^{-iky} +A_{k,s_i}e^{iky}\right)\Theta(y) +
-B_{k,s_i}e^{-iky}\Theta(-y)
\eea
where $\Theta(y)$ is the Heaviside step function, and the + and - refer to
incoming states from the left and right respectively.
(I will ignore the bound states, as the initial states I will choose will not
contain any bound state components).
The boundary conditions at $y=0$ give
\bea
1+A&=&B\\
ik((1-A)-B)&=&2m\mu\sum_i(s^3_i) B
\eea
(where I have suppressed the subscripts on the coefficients)
The solution is 
\bea
B= {2ik\over 2ik+2m\mu\sum(s^3_i)}\\
A=-{2m\mu\sum(s^3_i)\over 2ik+2m\mu\sum(s^3_i)}
\eea
The phase shift of the reflected and transmitted waves depends both on the
momentum and the value of $s^3_i$. In our case of a delta function coupling,
the maximum phase shift of the reflected wave (as compared with a hard wall
reflection) is 90 degrees. If one had a more complicated form for the spin
dependent scattering potential, instead of simply a delta function, one could
have larger phase shifts. 

Now, let us assume that the initial state is two well separated Gaussian  packets for
the two particles travelling toward each other , and that the state of each of 
the spins is the $+{1\over 2}$ eigenstates of the
$S^1_i$ operators at times far in the past.  We have that the state in the far
distant past is given by, where $\lambda_1$ and $\lambda_2$ are assumed to
have support only  for very large values of $x_1$ and $-x_2$ and whose
states in the momentum representation have the two particles travelling toward
each other. I.e., the intial state is a product state of the form
$\lambda_1(x_1)\lambda_2(x_2)\prod_i\ket{ s^1_i}$
For simplicity let me take
\bea
\lambda_1={1\over \sqrt{\sigma\sqrt{2\pi}}}e^{-x_1^2\over 4\sigma^2} e^{ik_0 x_1}
\\
\lambda_2={1\over \sqrt{\sigma\sqrt{2\pi}}}e^{-x_2^2\over 4\sigma^2} e^{-ik_0 x_2}
\eea
where $\sigma^2(t)=\sigma_0^2 + imt$. In the distant past ($t<<0$) these
represent two gaussian wave packets, one for each particle, travelling toward
each other with momentum $k_0$. Expressing these in terms of the $Y,y$
coordinates we have 
\bea
\lambda_1\lambda_2=  {1\over \sigma\sqrt{2\pi}}e^{-{Y^2\over 2\sigma^2}} e^{-{y^2\over 8\sigma^2}+ik_0 y}
\eea

Thus, long before the scattering, the state is assumed to be the product state
of the $s^1_i=+{1\over 2}$ eigenstates of the spin, the center of mass
state
\bea
\psi_{in}(t,Y)= {1\over 2\pi\sqrt{\sigma\sqrt{4\pi}}}\int \sqrt{\sigma}e^{-{(\sigma^2+2i{t\over m}\over 2}K^2}  e^{iKY}
dK
\eea
and the difference state
\bea
\phi_{in}(t,y)={1\over 2\pi\sqrt{\sigma\sqrt{\pi}}}\int\sqrt{\sigma}
e^{-{{2\sigma^2(k-k_0)^2 +i{t\over 2m}k^2}}}  e^{iky}dk
\eea

Long after the scattering ($t\rightarrow\infty$), the total state is
\bea
\Psi=
\psi(t,Y){1\over 2^{N/2}\pi\sqrt{\sigma\sqrt{\pi}}}\sum_{\{s^3_i=\pm 1\}} (\int_0^\infty A_{k,s^3_i}
e^{-iky}+B_{k,s^3_i} e^{iky}) e^{-i2mk^2t} e^{-\sigma^2 (k-k_0)^2}
dk\prod_i\ket{s^3_ia)}
\eea
Let me assume that $k_0\sigma>>0$ so that only the behaviour near
$k_0$ need be considered.

Now, we want to look at the reduced density matrix for the two particles. The
centre of mass motion $\psi(Y,t)$ is unaffected by anything, and will simply
factor out of the equations. Thus we need only look at the density matrix for
the relative coordinate $y$. This is most easily done in the momentum
representation, so that 
\bea
\rho(k,k')= {1\over 2\pi{\sigma\sqrt{\pi}}}e^{i2m(k^2-k'^2)t}{1\over 2^N}\sum_{s^3_i} \left[A^*_{k',s^3_i} A_{k,s^3_i}
e^{i2m(k^2-k'^2)t}\Theta(-k)\Theta(-k')+ B^*_{k',s^3_i}
B_{k,s^3_i}\Theta(k')\Theta(k) 
\right.
\\
\left.+B^*_{k',s^3_i}A_{k,s^3_i}\Theta(k')\Theta(-k)+A^*_{k',s^3_i}B_{k,s^3_i}\Theta(-k)\Theta(k'))\right]
\eea
where the assumption that the $N$  spins are all in the $S^1_i$ eigenstates is contained
in the factor of $1\over 2^N$ and the equal sum over all of the $s^3_i$
eigenstates. 

Let us assume that the Gaussian is very narrow in momentum
space(${1\over\sigma}< 2\mu m$) , so that we
can take $k=k'=k_0$.
 The on-diagonal terms are 
\bea
\rho(-k_0,-k_0)\propto {1\over 2^N}\sum_{s^3_i} {(2m\mu)^2(\sum_i
s^3_i)^2\over (2m\mu(\sum_i
s^3_i)^2+k_0^2 )}\\
\rho(k_0,k_0)\propto {1\over 2^N}\sum_{s^3_i} {k_0^2\over
((2m\mu(\sum_i
s^3_i))^2+k_0^2 )}\\
\rho(-k_0,k_0)=\rho(k_0,-k_0)=0
\eea
The last comes about because these terms are proportional to 
\bea
\sum_{s^3_i} {\sum_i s^3_i\over (2m\mu\sum_i s^3_i )^2 +k_0^2}
\eea
and this term  flips sign if we take all
$s^3_i\rightarrow -s^3_i$. Since both the positive and negative signs have
equal probability under the assumption we have made about the spin states,
these cancel.
I.e., this reduced density matrix for the scattering loses its off-diagonal 
terms in the momentum representation when one reduces over the spin space. 

This system clearly conserves energy in the particle sector. Long before and
long after the
interaction, the spins all have zero eigenstate of energy, and the energy is all in the two
particles. The total energy is conserved. But while the initial state is a pure state, the final density
matrix for the particles on their own is not. It has a non-zero entropy. For a broader initial state, one
also gets decoherence between the various $k$ values as well caused by the
different scatterings by the various spin states. 

Note that because of the form of the interaction (which depends only on the
difference of the two particle coordinates) this system also trivially conserves
momentum.  

While the above example has a minimal loss of coherence (entropy increase of
$ln(2)$ if $k_0$ has value $2\mu m N)$ this arises because of the assumption of
a very narrow packet in phase space. If the packet has a broad width of the order of
$2\mu mN$ there will also be loss of coherence between the energy eigenstates.
Essentially the decoherence is caused by a rotation of the spin vector. Since
for a total spin of S, the spin states can be distinguished if they rotate by
about $\Delta S=1$. Thus the states of various $k_0$ for which the spin rotation differ
by unity will be decohered by the scattering.

I am not arguing that the internal states of the black hole are such spin
states. I am however arguing that they share with the these spin degrees of
freedom the feature that after the black hole has evaporated they have zero
energy. The energy is all outside the black hole-- initially in the energies of the
particles that create the black hole, and finally in the particles that are evaporated from the black
hole. As mentioned, they could be in some sort of baby universe which split
off from the interior of the black hole. or they could simply represent the
degrees of freedom which fell off the edge of spacetime at the singularity
within the black hole. Gravity, unlike flat spacetime in which the intuition
of most  physicists was honed, has the feature that time can have an
end. Degrees of freedom can disappear by falling off the edge of the universe,
or they can appear from things like the initial singularity of the universe. 

Physicists are familiar with the loss of entropy in electromagnetic radiation
travelling to infinity, a loss which certainly does not entail any crisis for
physics. Now, that entropy is often associated with the loss of energy as
well, which is probably the origin of the intuition that there is some
relation between energy dissipation and entropy production. 
However, the necessity of that dissipation is simply not there. One can carry
off entropy in internal spin degrees of freedom. If we imagine that there
exists some particle with an internal spin of value $e^{10^{10}}$. Then while
the kinetic energy of that particle can be small, its entropy can be huge (up
to $10^{10}$). 

It is important to note that while those two particles are interacting with
each other, the total energy of the system is NOT completely contained in
those two particles. The presence of the particles at y=0 causes the spins to
precess, a precession which requires the spins to have energy. Ie, during the
interaction, the total energy is shared between the spins and the particles.
The spin-particle interaction-energy is proportional to $\sum\sigma^3_i$. While its expectation
value is zero in the chosen state, the expectation value of this operator
squared is non-zero. I.e., the interaction energy has  zero expectation value of energy
during the collision, but a large value of the energy squared. (a large
energy fluctuation). However, after the interaction has taken place, the
system is in an eigenstate 0 of the spin-particle interaction energy.   

This explicit counterexample to the claimed theorem of Benks {\it et al}
raises the question of where the problem in their proof occured. Their master
equation for the density matrix is precisely of the Lindblad
form\cite{lindblad}, an
equation which makes the Markovian assumption that the loss of coherence is
completely memoryless. The change in the density matrix is without memory, and
depends only on the current state. This of course means that it does not
remember how much energy has been emitted, and thus has no means of conserving
energy (part of the energy being hidden behind the horizon and inaccessible to
the outside world). While a reasonable approximation for many systems, the
above counterexample shows that it is not always a good approximation. 

The main lesson is that decoherence and energy need not be linked. Decoherence
can occur without energy loss. This is true of condensed matter systems and
possibly also for gravitational black-hole systems. Whether or not this
actually occurs for black holes is of course a fascinating and still open
question. But there is no crisis for physics if black holes do create loss of
coherence for the quantum fields outside the black hole.

\end{document}